\documentclass[preprint,aps,nofootinbib]{revtex4}
\usepackage{amsfonts,amsmath,amssymb,amsthm}
\usepackage{latexsym}
\usepackage{bbm,bm}
\usepackage{graphicx}

\newcommand{\ket}[1]{\lvert #1 \rangle}
\newcommand{\bra}[1]{\langle #1 \lvert}
\newcommand{\beq}{\begin{equation}}
\newcommand{\eeq}{\end{equation}}
\newcommand{\beqs}{\begin{eqnarray}}
\newcommand{\eeqs}{\end{eqnarray}}

\begin{document}

\title{Three-Tangle in Non-inertial Frame}

\author{Mi-Ra Hwang$^1$, Eylee Jung$^2$ and DaeKil Park$^{1,3}$}

\affiliation{$^1$Department of Electronic Engineering, Kyungnam University, ChangWon
              631-701, Korea                   \\
             $^2$Center for Superfunctional Materials,
                 Department of Chemistry,
                 Pohang University of Science and Technology, 
                 San 31, Hyojadong, Namgu, Pohang 790-784, Korea \\
             $^3$Department of Physics, Kyungnam University, ChangWon
                  631-701, Korea                                        
             }

\begin{abstract}
Let Alice, Bob, and Charlie initially share an arbitrary fermionic three-qubit pure state, whose three-tangle is $\tau_3^{(0)}$. 
It is shown within the single-mode approximation that if one party among the three 
of them moves with a uniform acceleration with respect to the other parties, the three-tangle reduces to 
$\tau_3^{(0)} \cos^2 r$, where $r$ denotes a statistical factor in Fermi-Dirac statistics.
\end{abstract}

\maketitle

Recently, quantum information theories in the relativistic framework have attracted considerable 
attention\cite{peres,inertial,non-inertial,beyond,single}. 
It seems to be mainly due to the fact that many modern experiments on quantum information processing involve the use of photons and/or electrons, where the relativistic effect is not negligible. Furthermore, relativistic quantum information is also important from purely
theoretical aspects\cite{theory} in the context of black hole physics and quantum gravity.

It has been shown in Ref.\cite{non-inertial} that the entanglement formed initially in an inertial frame is generally degraded in a non-inertial frame. 
In particular, the bipartite bosonic entanglement completely vanishes when one of the two parties approaches the Rindler horizon. 
However, for the tripartite case there is a remnant of bosonic entanglement even in the Rindler horizon. 

In this paper we will examine the degradation of the pure-state three-tangle in the non-inertial frame. We assume that the three 
stationary parties Alice, Bob, and Charlie share the arbitrary fermionic pure three-qubit state $\ket{\psi}$ 
at the event $P$ (see Fig. 1). After then, one
party undergoes constant acceleration while the remaining two parties remain at the rest frame. The worldlines for accelerating and 
stationary frames are described in Fig. 1 by red and blue solid lines, respectively. 

As Fig. 1 shows, the accelerating party cannot access the Rindler region $II$. This means that the accelerating party cannot communicate with 
any observer in this region. Thus, in order to describe physics from the point of view of the accelerating observer, we need to remove the part of 
the system described by the region $II$ due to its causally-disconnected nature. This can be achieved by taking a partial trace over this region, 
and, as a result, the information of the region $II$ is erased. The degradation of the bipartite entanglement\cite{non-inertial} occurs via this partial trace. 

One may argue that the partial trace over the region $II$ in the state of the accelerating party is not a physically reasonable process 
because the observers in the rest frame can communicate
with an observer in region $II$. This is true because the observer in region $II$ can send a signal to the observers in the rest frame 
when $t > \tau$. 
In spite of this fact the partial trace over $II$ for the accelerating party is a physical process due to the fact that the quantum entanglement 
is not a local property of the quantum state. Let us imagine that Alice, Bob, and Charlie have their own particle detectors.
Then, the entanglement of the quantum state can be measured through communication between rest and accelerating parties\footnote{This is one-way communication at $t > \tau$ because the observers in the rest frame are in the future wedge with respect to the accelerating party. Thus while the observer in the accelerating can send a signal to the observers in the rest frame, the reverse communication is impossible.}. 
Thus, the accelerating party send his (or her) data to the observers in the rest frame and the one of the observer in the rest frame can measure the quantity of the quantum entanglement by combining all data obtained from all particle detectors. In the data received from the accelerating party 
the effect of the partial trace over the region $II$ is already involved due to the causally-disconnected nature with the region $II$. Thus, 
the partial trace over the region $II$ is also physically realizable process from the aspects of the rest frame in the measurement of the quantum
entanglement.

The main result of this paper is that from the perspective of 
the accelerating parties the degradation factor of the three-tangle is independent of the initial state $\ket{\psi}$, but depends only on
the statistical factor. We show this in the following by making use of the single-mode approximation.

\begin{figure}[ht!]
\begin{center}
\includegraphics[height=10cm]{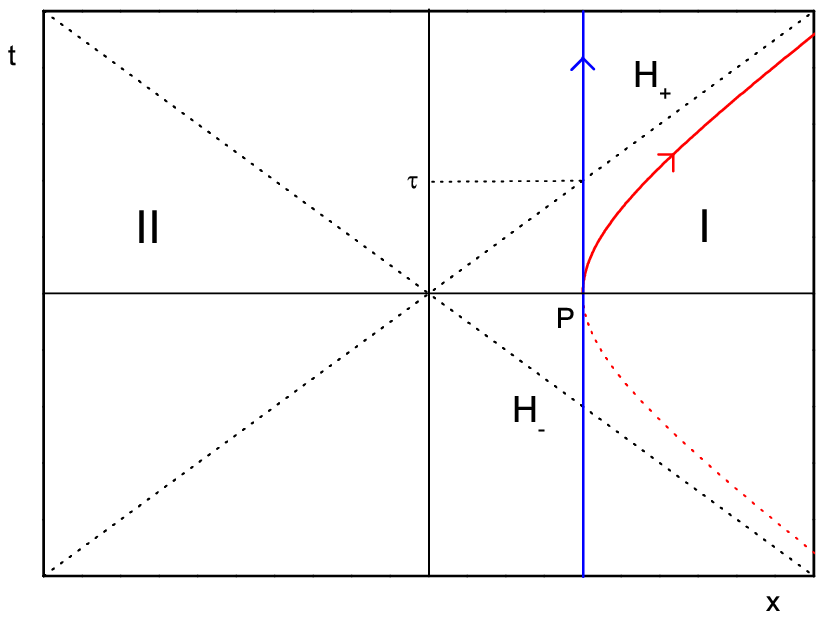}
\caption[fig1]{(Color online) We assume that the three parties, who are at rest frame at $t < 0$, share an arbitrary fermionic three-qubit 
pure state $\ket{\psi}$ at the event $P$. After then, one party undergoes a constant acceleration (red solid line) while the remaining two parties 
remain at the rest frame (blue solid line). The  $I$ and $II$ denote the two causally disconnected regions of Rindler space.}
\end{center}
\end{figure}

The fermionic entanglement can be addressed more easily than its bosonic counterpart due to the Pauli exclusion principle.
The Unruh decoherence beyond the single-mode approximation is fully discussed in Ref. \cite{beyond}. Let $\ket{n_{\Omega}}_R$ be 
a $n$-particle state with energy $E_{\Omega}$ in the spacetime $R$. Then, for the fermion case, the Unruh effect is described by 
\begin{eqnarray}
\label{not-single-1}
& &\ket{0_{\Omega}}_U = \ket{0_{\Omega}}_R \otimes \ket{0_{\Omega}}_L                                 \\   \nonumber
& &\ket{1_{\Omega}}_U^{+} = q_R \ket{1_{\Omega}}_R \otimes \ket{0_{\Omega}}_L + q_L \ket{0_{\Omega}}_R \otimes \ket{1_{\Omega}}_L,
\end{eqnarray}
where $|q_R|^2 + |q_L|^2 = 1$ and 
\begin{eqnarray}
\label{not-single-2}
& &\ket{0_{\Omega}}_R = \cos r_{\Omega} \ket{0_{\Omega}}_I^+ \ket{0_{\Omega}}_{II}^- + \sin r_{\Omega} \ket{1_{\Omega}}_I^+ \ket{1_{\Omega}}_{II}^-
                                                                                                                                    \\   \nonumber
& &\ket{0_{\Omega}}_L = \cos r_{\Omega} \ket{0_{\Omega}}_I^- \ket{0_{\Omega}}_{II}^+ - \sin r_{\Omega} \ket{1_{\Omega}}_I^- \ket{1_{\Omega}}_{II}^+
                                                                                                                                    \\   \nonumber
& &\ket{1_{\Omega}}_R =  \ket{1_{\Omega}}_I^+  \ket{0_{\Omega}}_{II}^-                    \\   \nonumber
& &\ket{1_{\Omega}}_L = \ket{0_{\Omega}}_I^- \ket{1_{\Omega}}_{II}^+.
\end{eqnarray}
Here, $\ket{n}_U$ is a $n$-particle state in Unruh mode and the $\pm$ indicates particle and antiparticle. The parameter $r_{\Omega}$ is 
a constant related to the Fermi-Dirac statistics. Since the Bogoliubov coefficients
$\beta_{jk}$ between the Unruh and Monkowski modes are zero, both modes share the common vacuum. However, in this paper we will not adopt 
Eq. (\ref{not-single-1}) but the simpler single-mode approximation because of the following reasons. 
Most important reason is the fact that Eq. (\ref{not-single-1}) makes the 
dimension of the Hilbert space for the accelerating party to be larger than the corresponding number for the stationary parties
due to particle and antiparticle modes. Since, so far, we know how to compute the three-tangle for only few low-rank qubit systems\cite{tangle}, 
the increase of the dimension generates a crucial difficulty for the analytical treatment of the three-tangle. However, the single mode
approximation does not generate these difficulties. Second reason is that as far as we know, 
the treatment of the three-tangle on the analytical ground in the non-inertial perspective is not discussed in other literature yet.
Thus, it is important to gain an insight in this issue. We think we can get a sufficient insight even though the single-mode approximation 
is adopted. Another reason is that the entanglement degradation with the single mode approximation is exactly the 
same with the degradation with $q_R = 1$ case\cite{beyond}. Thus, we can understand, at least, the degradation of the three-tangle in
the non-inertial frame for the special choice of Unruh mode. 
 
By computing the Bogoliubov coefficients\cite{Unruh, single} explicitly in the fermionic system,
one can show that within the single mode approximation, the vacuum state $\ket{0}_U$ and the one-particle state $\ket{1}_U$, where the 
subscript $U$ stands for Unruh mode, in a uniformly accelerated frame with acceleration $a$ are transformed into
\begin{eqnarray}
\label{unruh-1}
& &\ket{0}_U \rightarrow \cos r \ket{0}_I \ket{0}_{II} + \sin r \ket{1}_I \ket{1}_{II}             \\    \nonumber
& &\ket{1}_U \rightarrow \ket{1}_I \ket{0}_{II},
\end{eqnarray}
where the parameter $r$ is defined by 
\begin{equation}
\label{unruh-2}
\cos r = \frac{1}{\sqrt{1 + \exp (-2 \pi \omega c / a)}},
\end{equation}
and $c$ denotes the speed of light, and $\omega$ represents the central frequency of the fermion wave packet. It is to be noted 
that the sign in the denominator 
of Eq. (\ref{unruh-2}) is originated from Fermi-Dirac statistics. In Eq. (\ref{unruh-1}) $\ket{n}_I$ and $\ket{n}_{II}$ ($n=0, 1$) indicate the mode decomposition in two causally disconnected regions in Rindler space.

In this paper we will examine the three-tangle\cite{ckw}, one of the most important tripartite entanglement measures, in a non-inertial frame.
For this purpose we assume that Alice, Bob, and Charlie initially share an arbitrary pure three-qubit state $\ket{\psi}_{ABC}$. Applying a 
Schmidt decomposition we can transform $\ket{\psi}_{ABC}$ into the following\cite{acin00};
\begin{equation}
\label{schmidt}
\ket{\psi}_{ABC} = \lambda_0 \ket{000} + \lambda_1 e^{i \varphi} \ket{100} + \lambda_2 \ket{101} + \lambda_3 \ket{110} + \lambda_4 \ket{111}
\end{equation}
where $\lambda_i \geq 0 \hspace{.3cm} (i = 0, \cdots, 4)$, $\sum_i \lambda_i^2 = 1$, and $0 \leq \varphi \leq \pi$. The three-tangle of 
$\ket{\psi}_{ABC}$ is $\tau_3^{(0)} = 4 \lambda_0^2 \lambda_4^2$. We will show in our study that if one party moves with a uniform 
acceleration $a$ with respect to the other parties, the resulting three-tangle is degraded to 
\begin{equation}
\label{final}
\tau_3 = \tau_3^{(0)} \cos^2 r,
\end{equation}
regardless of which of the parties is accelerating. Therefore, at the Rindler horizon, the three-tangle reduces to half of the initial three-tangle.

Now, we assume that Alice, Bob, and Charlie were initially in the spacetime region $I$. If Alice was chosen as the accelerating party, the Unruh effect
(\ref{unruh-1}) transforms $\ket{\psi}_{ABC}$ into a four-qubit state $\ket{\psi_A}_{IBC\otimes II}$, whose explicit expression is 
\begin{eqnarray}
\label{alice-1}
& &\ket{\psi_A}_{IBC\otimes II} = \bigg[ \lambda_0 \cos r \ket{000} + \lambda_1 e^{i \varphi} \ket{100} + \lambda_2 \ket{101} + \lambda_3 \ket{110}
                                      + \lambda_4 \ket{111} \bigg] \otimes \ket{0}_{II}                 \\   \nonumber
& &\hspace{6.0cm}
                                      + \lambda_0 \sin r \ket{100}_{IBC} \otimes \ket{1}_{II}.
\end{eqnarray}
Since the accelerating party cannot access region $II$ due to the causally disconnected condition, it is reasonable to take a partial trace over $II$ to 
average the effect of qubit $\ket{n}_{II}$ out in Eq. (\ref{alice-1}). 
As a result, the initial state $\ket{\psi}_{ABC}$ reduces to a mixed state $\rho_{IBC} = \mbox{tr}_{II} \ket{\psi_A}\bra{\psi_A}$.
This means that some information formed initially in region $I$ is leaked into region $II$. Thus, Alice's acceleration induces an information loss,
which is a main consequence of the Unruh effect. 

One can show that the final state $\rho_{IBC}$ is a rank-$2$ tensor in a form
\begin{equation}
\label{alice-2}
\rho_{IBC} = p \ket{a_+}\bra{a_+} + (1 - p) \ket{a_-}\bra{a_-}
\end{equation}
where $p = (1 + \sqrt{\Delta}) / 2$ with 
\begin{equation}
\label{alice-3}
\Delta = \left(1 - 2 \lambda_0^2 \sin^2 r \right)^2 + 4 \lambda_0^2 \lambda_1^2 \sin^2 r.
\end{equation}
In Eq. (\ref{alice-2}) the vectors $\ket{a_{\pm}}$ are given by
\begin{equation}
\label{alice-4}
\ket{a_{\pm}} = \frac{1}{{\cal N}_{\pm}}
\bigg[ \lambda_0 \cos r z_{\pm} \ket{000} + y_{\pm} \ket{100} + \lambda_2 z_{\pm} \ket{101} + \lambda_3 z_{\pm} \ket{110} + \lambda_4 z_{\pm} \ket{111}
                \bigg]
\end{equation}
where
\begin{eqnarray}
\label{alice-5}
& & z_{\pm} = 1 - 2 \lambda_0^2 \sin^2 r \pm \sqrt{\Delta}                                                 \\   \nonumber
& &y_{\pm} =  e^{i \varphi} \lambda_1 (1 \pm \sqrt{\Delta})                                                   \\   \nonumber
& &{\cal N}_{\pm}^2 = (1 - \lambda_0^2 \sin^2 r - \lambda_1^2 ) z_{\pm}^2 + |y_{\pm}|^2.
\end{eqnarray}
It is easy to show $\bra{a_+}a_-\rangle = 0$, which guarantees that $\rho_{IBC}$ is a quantum state.

Since the three-tangle for mixed states is defined via a convex roof method\cite{convex-roof}, we should find an optimal decomposition of 
$\rho_{IBC}$ for analytical computation of the three-tangle. It is to be noted that the three-tangles for $\ket{a_{\pm}}$ are 
\begin{equation}
\label{alice-6}
\tau_3 \left( \ket{a_{\pm}} \right) = 4 \lambda_0^2 \lambda_4^2 \cos^2 r \left( \frac{z_{\pm}}{{\cal N}_{\pm}} \right)^4.
\end{equation}
Now, we define 
\begin{equation}
\label{alice-7}
\ket{F, \theta} = \sqrt{p} \ket{a_+} + e^{i \theta} \sqrt{1 - p} \ket{a_-}.
\end{equation}
Then, it is easy to show that $\rho_{IBC}$ can be represented as 
\begin{equation}
\label{alice-8}
\rho_{IBC} = \frac{1}{2} \ket{F,\theta}\bra{F,\theta} + \frac{1}{2} \ket{F,\theta+\pi}\bra{F,\theta+\pi}
\end{equation}
and the three-tangle of $\ket{F,\theta}$ is 
\begin{equation}
\label{alice-9}
\tau_3 \left(\ket{F, \theta} \right) = 4 \lambda_0^2 \lambda_4^2 \cos^2 r 
\left[\frac{p z_+^2}{{\cal N}_+^2} + \frac{(1-p) z_-^2}{{\cal N}_-^2} + \frac{2 \sqrt{p (1-p)} z_+ z_-}{{\cal N}_+ {\cal N}_-} \cos \theta \right]^2.
\end{equation}
Now, we assume that Eq. (\ref{alice-8}) is an optimal decomposition for the three-tangle for the time being. Then, the three-tangle of 
$\rho_{IBC}$ is 
\begin{equation}
\label{alice-10}
\tau_3 \left(\rho_{IBC} \right) = 4 \lambda_0^2 \lambda_4^2 \cos^2 r 
\left[ \left( \frac{p z_+^2}{{\cal N}_+^2} + \frac{(1-p) z_-^2}{{\cal N}_-^2} \right)^2 + 
       \frac{4 p (1-p) z_+^2 z_-^2}{{\cal N}_+^2 {\cal N}_-^2} \cos^2 \theta  \right].
\end{equation}
Therefore, the convex-roof constraint of the three-tangle leads to $\theta$ being fixed as $\theta = \pi/2$. 
Consequently, Eq. (\ref{alice-10}) indicates that $\tau_3 \left(\rho_{IBC} \right)$
is really a convex function with respect to $p$. Therefore, Eq. (\ref{alice-8}) is really an optimal decomposition of $\rho_{IBC}$ if 
$\theta = \pi/2$. By inserting Eq. (\ref{alice-3}) and Eq. (\ref{alice-5}) into Eq. (\ref{alice-10}) and imposing $\theta = \pi/2$, it is 
straight to show that
\begin{equation}
\label{alice-11}
\tau_3 \left(\rho_{IBC} \right) = 4 \lambda_0^2 \lambda_4^2 \cos^2 r.
\end{equation}
Thus, Eq. (\ref{final}) holds when Alice is chosen as the accelerating party.

Now, we choose Bob as the accelerating party. Following a procedure similar to the one described above, one can show that $\rho_{AIC}$ becomes
\begin{equation}
\label{bob-1}
\rho_{AIC} = p \ket{b_+}\bra{b_+} + (1 - p) \ket{b_-}\bra{b_-}
\end{equation}
where $p = (1 + \sqrt{\sigma}) / 2$ with 
\begin{equation}
\label{bob-2}
\sigma = \left[1 - 2 (\lambda_0^2 + \lambda_1^2 + \lambda_2^2 )\sin^2 r \right]^2 + 
         4 \sin^2 r \left(\lambda_1^2 \lambda_3^2 + \lambda_2^2 \lambda_4^2 + 2 \lambda_1 \lambda_2 \lambda_3 \lambda_4 \cos \varphi \right).
\end{equation}
In Eq. (\ref{bob-1}) the vectors $\ket{b_{\pm}}$ are given by
\begin{equation}
\label{bob-3}
\ket{b_{\pm}} = \frac{1}{{\cal N}_{\pm}} \left[ a_{000}^{\pm} \ket{000} + a_{010} \ket{010} + a_{100}^{\pm} \ket{100}
 + a_{101}^{\pm} \ket{101} + a_{110}^{\pm} \ket{110} + a_{111}^{\pm} \ket{111}  \right]
\end{equation}
where
\begin{eqnarray}
\label{bob-4}
& &a_{000}^{\pm} = \lambda_0 \cos r \left[ 1 - 2 \sin^2 r (\lambda_0^2 + \lambda_1^2 + \lambda_2^2 ) \pm \sqrt{\sigma} \right]    
                                                                                                          \\    \nonumber
& &a_{010} = 2 \lambda_0 \sin^2 r \left[e^{-i \varphi} \lambda_1 \lambda_3 + \lambda_2 \lambda_4 \right]   \\    \nonumber
& &a_{100}^{\pm} = e^{i \varphi} \lambda_1 \cos r \left[ 1 - 2 \sin^2 r (\lambda_0^2 + \lambda_1^2 + \lambda_2^2 ) \pm \sqrt{\sigma} \right]    
                                                                                                          \\    \nonumber
& &a_{101}^{\pm} = \lambda_2 \cos r \left[ 1 - 2 \sin^2 r (\lambda_0^2 + \lambda_1^2 + \lambda_2^2 ) \pm \sqrt{\sigma} \right]    
                                                                                                          \\    \nonumber
& &a_{110}^{\pm} = \lambda_3 \left[ 1 - 2 \sin^2 r (\lambda_0^2 + \lambda_2^2 ) \pm \sqrt{\sigma} \right] + 
                   2 e^{i \varphi} \lambda_1 \lambda_2 \lambda_4 \sin^2 r                                     \\    \nonumber
& &a_{111}^{\pm} = \lambda_4 \left[ 1 - 2 \sin^2 r (\lambda_0^2 + \lambda_1^2 ) \pm \sqrt{\sigma} \right] + 
                   2 e^{-i \varphi} \lambda_1 \lambda_2 \lambda_3 \sin^2 r
\end{eqnarray}
and ${\cal N}_{\pm}^2 = |a_{000}^{\pm}|^2 + |a_{010}|^2 + |a_{100}^{\pm}|^2 + |a_{101}^{\pm}|^2 + |a_{110}^{\pm}|^2 + |a_{111}^{\pm}|^2$. It is 
easy to show that the three-tangles of $\ket{b_{\pm}}$ are
\begin{equation}
\label{bob-5}
\tau_3 \left( \ket{b_{\pm}} \right) = 4 \lambda_0^2 \lambda_4^2 \cos^2 r 
\left[ \frac{1 - 2 \sin^2 r (\lambda_0^2 + \lambda_1^2 + \lambda_2^2 ) \pm \sqrt{\sigma}}{{\cal N}_{\pm}} \right]^4.
\end{equation}
In order to compute the three-tangle of $\rho_{AIC}$ we define
\begin{equation}
\label{bob-6}
\ket{G,\theta} = \sqrt{p} \ket{b_+} + e^{i \theta} \sqrt{1-p} \ket{b_-}.
\end{equation}
Then, $\rho_{AIC}$ can be represented as 
\begin{equation}
\label{bob-7}
\rho_{AIC} = \frac{1}{2} \ket{G,\theta} \bra{G,\theta} + \frac{1}{2} \ket{G,\theta+\pi} \bra{G,\theta+\pi}
\end{equation}
and the three-tangle of $\ket{G,\theta}$ is 
\begin{equation}
\label{bob-8}
\tau_3 \left( \ket{G, \theta} \right) = 
4 \lambda_0^2 \lambda_4^2 \cos^2 r \left[ (X - Y)^2 + Z^2 + 4 X Y \cos^2 \theta -2 Z (X + Y) \cos \theta \right]
\end{equation}
where
\begin{eqnarray}
\label{bob-9}
& &X = \frac{p}{{\cal N}_+^2} \left[1 - 2 \sin^2 r (\lambda_0^2 + \lambda_1^2 + \lambda_2^2 ) + \sqrt{\sigma} \right]^2
                                                                                                                        \\   \nonumber
& &Y = \frac{1 - p}{{\cal N}_-^2} \left[1 - 2 \sin^2 r (\lambda_0^2 + \lambda_1^2 + \lambda_2^2 ) - \sqrt{\sigma} \right]^2
                                                                                                                        \\   \nonumber
& &Z = 8 \sin^2 r \frac{\sqrt{p (1 - p)}}{{\cal N}_+ {\cal N}_-} 
       \left[ \lambda_1^2 \lambda_3^2 + \lambda_2^2 \lambda_4^2 + 2 \lambda_1 \lambda_2 \lambda_3 \lambda_4 \cos \varphi \right].
\end{eqnarray}
Therefore, if Eq. (\ref{bob-7}) is the optimal decomposition, the three-tangle of $\rho_{AIC}$ is 
\begin{equation}
\label{bob-10}
\tau_3 \left( \rho_{AIC} \right)
= 4 \lambda_0^2 \lambda_4^2 \cos^2 r \left[  (X - Y)^2 + Z^2 + 4 X Y \cos^2 \theta \right].
\end{equation}
Since the three-tangle is defined as a convex roof method, we should choose $\theta = \pi / 2$, which gives
\begin{equation}
\label{bob-11}
\tau_3 \left( \rho_{AIC} \right)
= 4 \lambda_0^2 \lambda_4^2 \cos^2 r \left[  (X - Y)^2 + Z^2 \right].
\end{equation}
It is easy to show that Eq. (\ref{bob-11}) is really a convex function with respect to $p$. Using of Eq. (\ref{bob-2}), Eq. (\ref{bob-4}), 
and Eq. (\ref{bob-9}) and performing a series of calculations, we can show that $(X- Y)^2 + Z^2 = 1$, which results in 
\begin{equation}
\label{bob-12}
\tau_3 \left( \rho_{AIC} \right) = 4 \lambda_0^2 \lambda_4^2 \cos^2 r.
\end{equation}
Thus, Eq. (\ref{final}) holds when Bob is chosen as the accelerating party.

Finally, let us choose Charlie as the accelerating party.
Since $\ket{\psi}_{ABC}$ given in Eq. (\ref{schmidt}) has  Bob $\leftrightarrow$ Charlie and $\lambda_2 \leftrightarrow \lambda_3$ symmetry,
the previous calculation implies
\begin{equation}
\label{charlie-1}
\tau_3 (\rho_{ABI}) = 4 \lambda_0^2 \lambda_4^2 \cos^2 r.
\end{equation}
Thus, Eq. (\ref{final}) holds regardless of the choice of the accelerating party.

So far, we have shown that the canonical form of the three-qubit state (\ref{schmidt}) obeys Eq. (\ref{final}) in the non-inertial frame. However, 
this does not mean that the arbitrary three-qubit pure state $\ket{\psi_3} = \sum_{i,j,k=0}^1 a_{ijk} \ket{ijk}$ obeys Eq. (\ref{final}). 
The question arises as to whether the Unruh transformation (\ref{unruh-1}) that is taken before the appropriate Schmidt decomposition may generate a result different from  
from Eq. (\ref{final}). However, this is not the case; the following two theorems show that $\ket{\psi_3}$ also obeys Eq. (\ref{final}) in 
the non-inertial frame.

{\bf Theorem 1.} {\it Let Alice and Bob initially share the arbitrary fermionic two-qubit pure state 
$\ket{\psi_2}_{AB} = \sum_{i,j=0}^1 a_{ij} \ket{ij}$,
whose concurrence is $\tau_2^{(0)}$. If one party accelerates with respect to the other party, then the concurrence reduces to $\tau_2^{(0)} \cos r$.} 

\smallskip

{\bf Proof.} It is to be noted that the concurrence of $\ket{\psi_2}_{AB}$ is 
\begin{equation}
\label{th-1}
\tau_2^{(0)} = 2 | a_{00} a_{11} - a_{01} a_{10} |.
\end{equation}
First, we choose Bob as the accelerating party. Then, the Unruh transformation (\ref{unruh-1}) and the partial trace over $II$ gives
\begin{eqnarray}
\label{th-2}
\rho_{AI} = \left(                    \begin{array}{cccc}
                  |a_{00}|^2 \cos^2 r & a_{00} a_{01}^* \cos r & a_{00} a_{10}^* \cos^2 r & a_{00}a_{11}^* \cos r            \\
                  a_{00}^* a_{01} \cos r & |a_{01}|^2 + |a_{00}|^2 \sin^2 r & a_{01} a_{10}^* \cos r & a_{01} a_{11}^* + a_{00} a_{10}^* \sin^2 r \\
                  a_{00}^* a_{10} \cos^2 r & a_{01}^* a_{10} \cos r & |a_{10}|^2 \cos^2 r & a_{10} a_{11}^* \cos r           \\
                  a_{00}^* a_{11} \cos r & a_{01}^* a_{11} + a_{00}^* a_{10} \sin^2 r & a_{10}^* a_{11} \cos r & |a_{11}|^2 + |a_{10}|^2 \sin^2 r
                                       \end{array}                        \right).
\end{eqnarray}
Using Eq. (\ref{th-2}) we can construct $R = \rho_{AI} (\sigma_y \otimes \sigma_y) \rho_{AI}^* (\sigma_y \otimes \sigma_y)$. Although $R$ is a
complicated matrix, it is possible to compute the eigenvalues analytically by solving $\mbox{det} (R - \lambda I) = 0$. The eigenvalues of $R$ are 
$\left\{0, 0, 0, 4 |a_{00} a_{11} - a_{01} a_{10} |^2 \cos^2 r \right\}$. Therefore, by making use of the Wootters formula\cite{wootters}, 
we obtain the concurrence of $\rho_{AI}$ as 
\begin{equation}
\label{th-3}
\tau_2 (\rho_{AI}) = 2 | a_{00} a_{11} - a_{01} a_{10} | \cos r = \tau_2^{(0)} \cos r.
\end{equation}
Since $\tau_2^{(0)}$ has $a_{01} \leftrightarrow a_{10}$ symmetry, the choice of Alice as the accelerating party leads to the same result, 
which completes the proof.

Now, we state the main theorem of the paper.

{\bf Theorem 2.} {\it Let Alice, Bob, and Charlie initially share the arbitrary fermionic three-qubit pure state 
$\ket{\psi_3}_{ABC} = \sum_{i,j,k=0}^1 a_{ijk} \ket{ijk}$,
whose three-tangle is $\tau_3^{(0)}$. If one party accelerates with respect to other parties, then the three-tangle reduces to $\tau_3^{(0)} \cos^2 r$.} 

\smallskip

{\bf Proof.} It is to be noted that the three-tangle of $\ket{\psi_3}$\cite{ckw} is 
\begin{equation}
\label{th-4}
\tau_3^{(0)} = 4 |d_1 - 2 d_2 + 4 d_3|,
\end{equation}
where
\begin{eqnarray}
\label{th-5}
& &d_1 = a^2_{000} a^2_{111} + a^2_{001} a^2_{110} + a^2_{010} a^2_{101} + 
                                                              a^2_{100} a^2_{011},
                                                              \\   \nonumber
& &d_2 = a_{000} a_{111} a_{011} a_{100} + a_{000} a_{111} a_{101} a_{010} + 
         a_{000} a_{111} a_{110} a_{001}
                                                              \\   \nonumber
& &\hspace{1.0cm} +
         a_{011} a_{100} a_{101} a_{010} + a_{011} a_{100} a_{110} a_{001} + 
         a_{101} a_{010} a_{110} a_{001},
                                                              \\   \nonumber
& &d_3 = a_{000} a_{110} a_{101} a_{011} + a_{111} a_{001} a_{010} a_{100}. 
\end{eqnarray}
Since $\tau_3^{(0)}$ is permutation-invariant, it is sufficient to provide proof for the case that Alice is chosen as the 
accelerating party. We provide the proof via two different methods. The first proof is simple but intuitive while the second is lengthy 
and direct. Thus, we present the second one schematically.

As shown in Ref. \cite{ckw} the three-tangle is defined via the monogamy inequality ${\cal C}_{AB}^2 + {\cal C}_{AC}^2 \leq {\cal C}_{A(BC)}^2$,
where ${\cal C} = \tau_2$ is a concurrence. Therefore, the three-tangle of the mixed state can be written as 
\begin{equation}
\label{th-6}
\tau_3 = \min \left[ {\cal C}^2_{A(BC)} - {\cal C}^2_{AB} - {\cal C}^2_{AC} \right],
\end{equation}
where the minimum is taken over all possible decompositions of the given mixed state. Since Alice is chosen as the accelerating party, 
theorem $1$ implies 
that each concurrence in Eq. (\ref{th-6}) has a degradation factor $\cos r$. Therefore, Eq. (\ref{th-6}) implies that the three-tangle 
has a degradation factor $\cos^2 r$, which is what theorem $2$ states.

Another method to prove theorem $2$ is similar to the method for the proof of theorem $1$. After performing the Unruh transformation (\ref{unruh-1}) on 
Alice's qubit of $\ket{\psi_3}_{ABC}$ and taking a partial trace over $II$, one can derive $\rho_{IBC}$ straightforwardly. Although 
$\rho_{IBC}$ is an extremely complicated $8 \times 8$ matrix, one can show from a purification protocol that its rank is only $2$. 
Therefore, it is possible to derive the spectral 
decomposition of $\rho_{IBC}$ as a form $\rho_{IBC} = p \ket{\mu_+}\bra{\mu_+} + (1-p) \ket{\mu_-}\bra{\mu_-}$. 
Consequently, following a procedure similar to the one we used previously, we can compute the three-tangle of $\rho_{IBC}$ explicitly. 
We perform this calculation 
by making use of the software Mathematica, and we finally arrive at Eq. (\ref{final}) again, which completes the proof.

In this paper we investigated the degradation of the tripartite fermionic entanglement in a non-inertial frame. If the given three parties 
initially share a 
pure state, the degradation factor is shown to be simply $\cos^2 r$ regardless of the initial state and choice of the accelerating party. This is a 
surprising result in the sense of the simpleness of Eq. (\ref{final}). 

It is natural to ask whether or not the simpleness of Eq. (\ref{final}) is maintained when the initial state is a mixed state. 
Another natural question is to ask whether or not the simpleness of Eq. (\ref{final}) is maintained beyond the single mode approximation. 
As commented earlier Eq. (\ref{not-single-1}) generates the increase of the Hilbert space dimension for accelerating party. This gives a difficulties 
for the computation of the three-tangle. So far, we do not know how to define the three-tangle even in the qudit system.  
We believe that if the calculation tool for the three-tangle in the higher-dimensional Hilbert space is developed, 
Eq. (\ref{final}) or other such simple expressions are valid in these cases. 

In order to escape the increase of the Hilbert space dimension, it is possible to consider the particle or antiparticle sector in the 
full Hilbert space of the accelerating
party in Eq. (\ref{not-single-1}). In this case, however, one has to trace over multiqubit states in the whole density matrices, which generally
makes the rank of the final state larger than two. For example, if we consider the Charlie's particle sector of $\ket{\psi}_{ABC}$ given in 
Eq. (\ref{schmidt}), we have to trace over $I^-$, 
$II^+$, and $II^-$ in the density matrices and the final state reduces to 
\begin{eqnarray}
\label{ABIp}
& &\rho_{ABI^+} = \mbox{Tr}_{I^-, II^+, II^-} \rho                                                      \\    \nonumber
& &= \left(           \begin{array}{cccccccc}
\lambda_0^2 C^2 & 0 & 0 & 0 & \lambda_0 \lambda_1 e^{-i \varphi} C^2 & q_R^* \lambda_0 \lambda_2 C & \lambda_0 \lambda_3 C^2 & q_R^* \lambda_0\lambda_4 C
                                                                                                                                               \\
0 & \lambda_0^2 S^2 & 0 & 0 & 0 & \lambda_0 \lambda_1 e^{-i \varphi} S^2 & 0 & \lambda_0 \lambda_3 S^2                                         \\
0 & 0 & 0 & 0 & 0 & 0 & 0 & 0                                                                                                                  \\
0 & 0 & 0 & 0 & 0 & 0 & 0 & 0                                                                                                                   \\
\lambda_0 \lambda_1 e^{i \varphi} C^2 & 0 & 0 & 0 & T_1 C^2 & q_R^* \lambda_1 \lambda_2 e^{i\varphi} C &T_2 C^2 & q_R^*\lambda_1\lambda_4 e^{i \varphi} C
                                                                                                                                                 \\
q_R \lambda_0 \lambda_2 C & \lambda_0 \lambda_1 e^{i \varphi} S^2 & 0 & 0 & q_R \lambda_1 \lambda_2 e^{-i\varphi} C & T_3 S^2 + T_4 C^2 & 
q_R \lambda_2 \lambda_3 C & T_5 S^2 + T_6 C^2                                                                                                     \\
\lambda_0 \lambda_3 C^2 & 0 & 0 & 0 & T_2^* C^2 & q_R^* \lambda_2 \lambda_3 C & T_7 C^2 & q_R^* \lambda_3 \lambda_4 C                                \\
q_R \lambda_0 \lambda_4 C & \lambda_0 \lambda_3 S^2 & 0 & 0 & q_R \lambda_1\lambda_4 e^{-i \varphi} C & T_5^* S^2 + T_6 C^2 
& q_R \lambda_3 \lambda_4 C & T_8 S^2 + T_9 C^2
                                 \end{array}                                              \right),
\end{eqnarray}
where $C = \cos r_{\Omega}$, $S = \sin r_{\Omega}$, $T_1 = \lambda_1^2 + |q_L|^2 \lambda_2^2$, 
$T_2 = \lambda_1 \lambda_3 e^{i \varphi} + |q_L|^2 \lambda_2 \lambda_4$, $T_3 = \lambda_1^2 + \lambda_2^2$, $T_4 = |q_R|^2 \lambda_2^2$,
$T_5 = \lambda_1 \lambda_3 e^{i \varphi} + \lambda_2 \lambda_4$, $T_6 = |q_R|^2 \lambda_2 \lambda_4$, $T_7 = \lambda_3^2 + |q_L|^2 \lambda_4^2$,
$T_8 = \lambda_3^2 + \lambda_4^2$, and $T_9 = |q_R|^2 \lambda_4^2$. It is easy to show that the rank of $\rho_{ABI^+}$ is four and, furthermore, the 
nonvanishing eigenvalues cannot be obtained analytically. Since the analytical computation of three-tangle for a higher-rank mixed state is 
generally impossible except very rare cases\cite{tangle}, we think it seems to be impossible to compute the three-tangle for 
$\rho_{ABI^+}$ analytically. Similar situation occurs in the antiparticle sector.

Another question that arises is the extension of Eq. (\ref{final}) to multipartite entanglement. 
If an $n$-tangle is constructed in the future, we speculate
that the degradation factor in a non-inertial frame would be $\cos^{\alpha_n} r$, where $\alpha_n = 2^{n-2}$. However, there are several 
obstacles to confirming this hypothesis. Above all, the explicit expression of $n$-tangle is not yet known. 
Furthermore, there is no calculational technique for the computation of the $n$-tangle of $n$-qubit mixed states.
Our future studies will further explore these issues.

{\bf Acknowledgement}:
This research was supported by the Basic Science Research Program through the National Research Foundation of Korea(NRF) funded by the Ministry of Education, Science and Technology(2011-0011971).

\end{document}